\documentclass[12pt,square]{iopart}
\usepackage{graphicx}
\usepackage{epstopdf}
\usepackage{dcolumn}
\usepackage{bm}
\usepackage{iopams}
\usepackage{multirow}
\usepackage{citesort}
\begin{document}
\title[The 3D thermal spike model applied to nanoparticle irradiation with swift heavy ions]{Ion-matter interaction: the three dimensional version of the thermal spike model. Application to nanoparticle irradiation with swift heavy ions}
\author{Ch Dufour $^1$, V Khomenkov$^1$, G Rizza$^2$, M Toulemonde $^1$ }
\ead{christian.dufour@ensicaen.fr}
\address{$^1$ CIMAP, CEA / CNRS / ENSICAEN / Universit\'e de Caen, 6 Boulevard du Mar\'echal Juin, 14050 Caen cedex 4, France}
\address{$^2$ Ecole Polytechnique, Laboratoire des Solides Irradi\'es (LSI) CEA/DSM/IRAMIS, CNRS 91128 Palaiseau Cedex, France  }
%
\date{\today}
\begin{abstract}
In the framework of swift heavy ion - matter interaction, the thermal spike has proved its worth since nearly two decades. This paper deals with the necessary refinement of the computation due to the kind of materials involved i.e. nanomaterials such as multilayered systems or composite films constitued of nanocylinders or nanospheres embedded in matrix. The three dimensional computation of the thermal spike model is applied for the first time in layers containing spherical nanoparticles embedded in a silica matrix. The temperature profile calculated at each point (x,y,z) of the target for times up to $10^{-10}$s allows a possible explanation of the particle shape change under irradiation with swift heavy ions having an energy of several MeV/u.m.a. The comparison made with the former 2D version of the code applied to cylindrical gold nanoparticles confirms the validity of the present 3D version.
\end{abstract}
\submitto{\JPD}
\noindent{\it Ion radiation effects, swift heavy ion, thermal spike, nanoparticle\/}
\maketitle

\section{\label{sec:level1}Metallic nanoparticles synthesis with ion beam size and shape}
Since a decade, materials based on nanoparticles within a dielectric matrix have been elaborated with various techniques among which the use of swift heavy ions \cite{DOrleans2003}. Such kind of materials have been found interesting because of their peculiar optical properties related to plasmon excitation by UV-visible radiation \cite{Singhal2009}. Potential applications are expected particularly in the field of optical sensors. It has been demonstrated that plasmon frequency depends on the particle size, on their shape (spherical, cylindrical) and also on the surrounding matrix \cite{Mishra2007,Kuiri2010}. It is then important to monitor such geometrical parameters with ion beam assisted method cited above. Authors use one-step \cite{Kumar2008,Mishra2007,Mishra2007b,Singh2009} or two-step processes \cite{Silva-Pereyra2010,DOrleans2003} which combine a prior ion implantation with the desired metallic species with	post elaboration irrad
 iation.
Since 2003 \cite{DOrleans2003}, several papers invoke the thermal spike process leading to the shaping of nanoparticles \cite{Rizza2007,Rizza2009,Dawi2009} and we particularly refer to the work of Ridgway et al.\cite{Ridgway2011}. We have been among the first teams to delevop numerically this model since $1992$ \cite{Toulemonde1992}. We have applied successfully this model in a wide range of materials: metals \cite{Dufourcondmatt1993,WANG1994,WANG1995}, semiconducting and insulators \cite{Toulemonde1996,Toulemonde2000}.
In this paper, we propose the first real 3D version of the model which proves necessary in cases where the usual cylindrical symmetry is no longer valid: anisotropic thermal parameters and/or anisotropy in the geometry of the physical problem. We first discuss the thermal parameters that need special attention in order to give quantitative information instead of qualitative description. This discussion is made in relation with the work of Awazu et al. \cite{Awazu2008} and gives light on the choice of the thermal parameters. We therefore perform our simulation for the materials and ions studied in \cite{Awazu2008}: cylindrical gold nanoparticles embedded in silica and irradiated with Br 110 MeV. Then we focus on the application of the model to the nanoparticle transformation under irradiation with swift heavy ions and show the difference between cylindrical and spherical particles such as those investigated in \cite{Rizza2007,Rizza2009,Dawi2009,Ridgway2011}. Finally we provide a 
 full description of the behaviour of a spherical particle due to an ion impact.

\section{\label{TS3D}The 3D thermal spike model}

The target material is considered as a two component system: the atomic lattice and the electrons characterized by their respective temperatures $T_e$ and $T_a$.
Starting from the set a coupled equations (\eref{equadiffa} and \eref{equadiffb}) governing the electronic and atomic temperatures ($T_e$ and $T_a$), every physical parameter may now depend on the position $\vec{r}(x,y,z)$ in space. We write the energy variation $dQ=CdT$ linked to the temperature variation $dT$ of an elementary volume $dV$ characterized by a specific heat $C$. $dQ$ is due to three terms: $i)$ the energy brought by the incident ion on the electrons ($A_e(\vec{r},t)$) and on the atomic lattice ($A_a(\vec{r},t)$), $ii)$ the Fick's thermal diffusion law which defines the heat flux $\vec{j}= \overline{\overline{K}}\cdot\nabla T$ where the thermal conductivity $\overline{\overline{K}}$ is a 3$\times$3 matrix with the components $K_{ij}$, and $iii)$ the energy exchange between electrons and atomic lattice proportional to the temperature difference  $g(\vec{r},t)(T_e-T_a)$ where $g(\vec{r},t)$ the electron phonon coupling constant.

\numparts
\begin{eqnarray}
C_e\frac {\partial T_e}{\partial t}= \nabla(K_e\cdot\nabla T_e) - g(T_e-T_a)+A_e \label{equadiffa}\\
C_a\frac {\partial T_a}{\partial t}= \nabla(\overline{\overline{K_a}}\cdot\nabla T_a) + g(T_e-T_a)+A_a \label{equadiffb}
\end{eqnarray}
\endnumparts

We describe the different terms in the following sections.

\subsection{Incident ion energy deposition}
The linear energy transfer from the incident ion to the target atoms ($S_n$) and electrons ($S_e$) were computed using SRIM code by Biersack et al.\cite{Ziegler2010} and shown in the Table \tref{tab:parameters}. In the case of swift ions $S_n / S_e < 10^{-2}$ so that we neglect the nuclear component $A_a$ and deal only with electronic one $A_e$.

\begin{figure}
   \begin{minipage}{0.5\textwidth}   
      \centering {\includegraphics[scale=0.3]{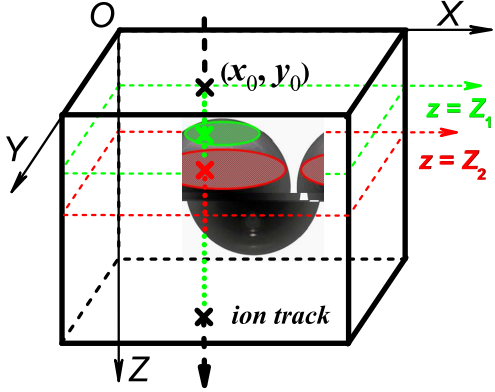}}
   \end{minipage}\\
 \caption{Geometry of the calculation box.}
 \label{figschema}
\end{figure}

We suppose that the incident ion hits the target parallel to $Z$-axis on $(x_0,y_0)$ point in $XY$-plane (\fref{figschema})
We separate the space and time variables so that $A_e(\vec r,t)=F(\vec{r})G(t)$. The function $G(t)$ is a time pulse of few femtoseconds duration ($t_0=2~ 10^{-15}$s), and is normalized so that $\int_0^{t_0}G(t) dt=1$. 

The function $F(\vec{r})$ is a product $F(\vec{r})=C~ S_e(x_0,y_0,z)~ \nu^{2}(x,y,z)~ W(r_{\bot})$. Here we introduce some new parameters:
$\nu(x,y,z)=\rho Z/A$ is the parameter proportional to the electron density ($\rho$ is mass density, $Z$ and $A$ are atomic and mass number of target material); 

$r_{\bot}(x,y,z)=\bar{\nu}~ \sqrt{(x-x_0)^2+(y-y_0)^2}$ is the radial range (in g$~$cm$^{-2}$) to the ion track, weighted by mean electron density; 

$\bar{\nu}=\int_{0}^{1}\nu\left[\xi(\vartheta),\eta(\vartheta),z\right]d\vartheta$ where $\xi(\vartheta)=x_0+(x-x_0)\vartheta$ and $\eta(\vartheta)=y_0+(y-y_0)\vartheta$.

Finally, $W(r_{\bot})= \frac{[1-(r_{\bot}+R_{min})/(R_{max}+R_{min})]^{1/\alpha}}{r_{\bot}~(r_{\bot}+R_{min})}$ is known as the radial $\delta$-ray energy distribution (Waligorski et al. \cite{WALIGORSKI1986}), where we substitute the range for homogeneous material with our weighted range $r_{\bot}$. According to \cite{WALIGORSKI1986}, $R_{min}$, $R_{max}$ are the ranges corresponding to the target ionization potential $E_{min}$ and maximum energy of $\delta$-electrons $E_{max}=2mc^2 \beta^2 /(1-\beta^2)$, $\beta$ is incident ion velocity in lightspeed units; $R=k ~ E^\alpha$, $k=6~10^{-6}$g$~ $cm$^{-2}~ $keV$^{-\alpha}$; $\alpha=1.079$ for $\beta<0.03$ and $\alpha=1.667$ for $\beta>0.03$;  $W(r_{\bot}>R_{max})=0$.

$C$ is a normalizing constant, so that $\int \int_{-\infty}^{\infty}F(x,y,z) dx dy=S_e(z)$.
\begin{figure}
  \begin{minipage}{0.5\textwidth}
      \centering {\includegraphics[scale=0.6]{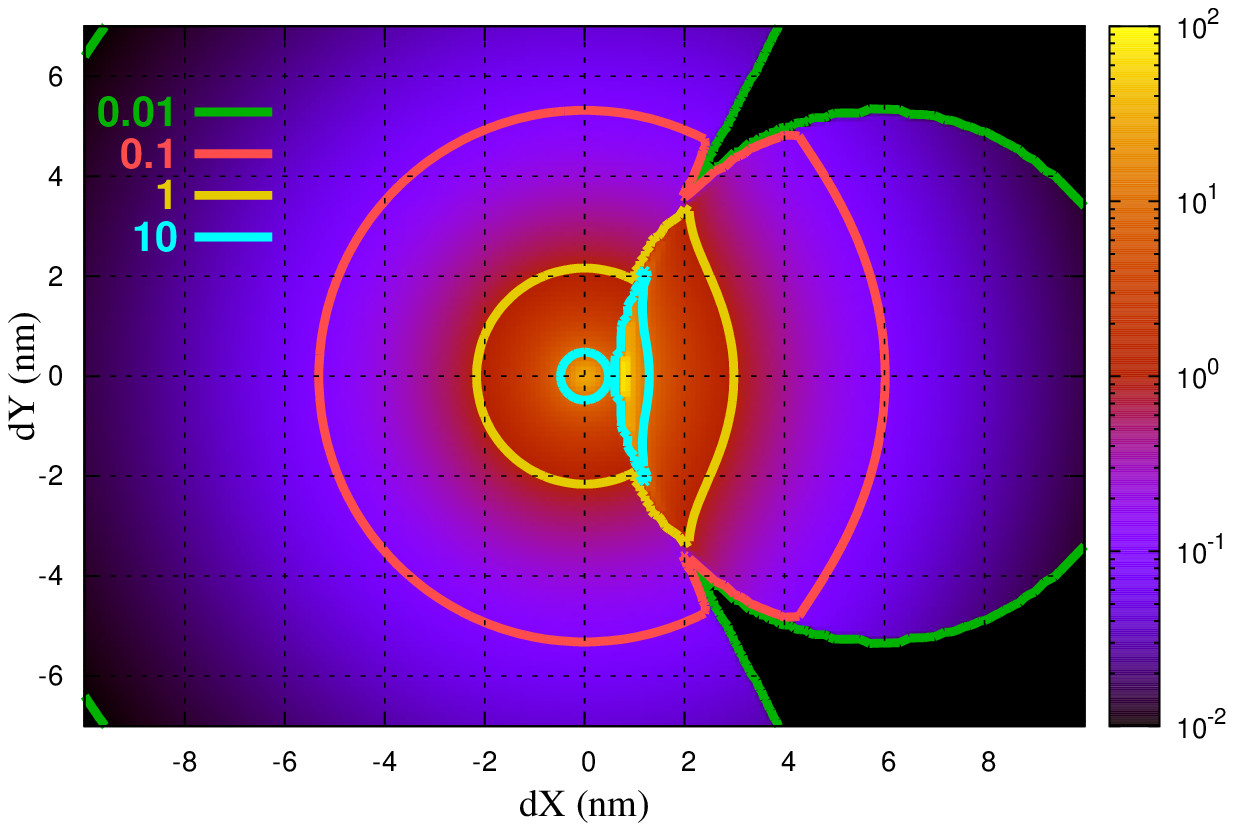}\\a)}
   \end{minipage}\hfill
   \begin{minipage}{0.5\textwidth}   
      \centering {\includegraphics[scale=0.6]{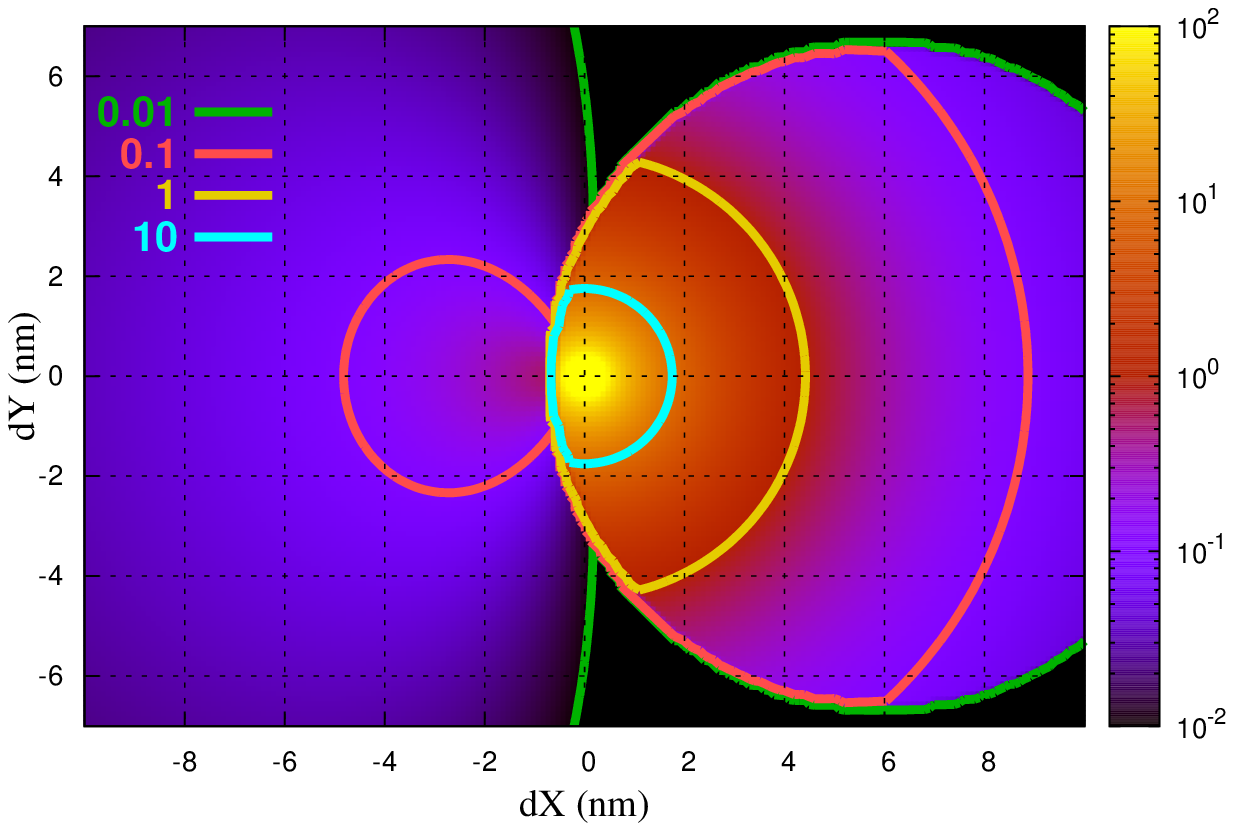}\\b)}
   \end{minipage}\\
 \caption{Initial energy distribution on the target electrons ($\delta$-electrons) [eV/atom] in different $XY$-planes for a) $Z=Z_1$ and b)$Z=Z_2$.}
 \label{WalXY}
\end{figure}

The energy density transferred to the $\delta$ electrons is plotted (in eV/atom) in two $XY$ planes for two values of $Z$ (\fref{WalXY}). Hence, we see the distributions in the upper plane ($Z=Z_1$) for which the ion track is outside the nanoparticle (\fref{WalXY} a)) and in the bottom plane ($Z=Z_2$) (\fref{WalXY} b)). Due to the different electronic densities of the nanoparticle and the matrix, we notice that the energy distribution is wider within the nanoparticle than in the matrix (e.g. see the light blue contour corresponding to a zone in which the energy density exceeds $10$ eV/atom). The different shapes of spatial energy distribution proves the influence of the nanoparticle / matrix interface at this first step of the ion-matter interaction. 

\subsection{Lattice parameters} \label{lattparam}
\begin{figure}
  \begin{minipage}{0.5\textwidth}
		\centering \includegraphics[scale=0.25]{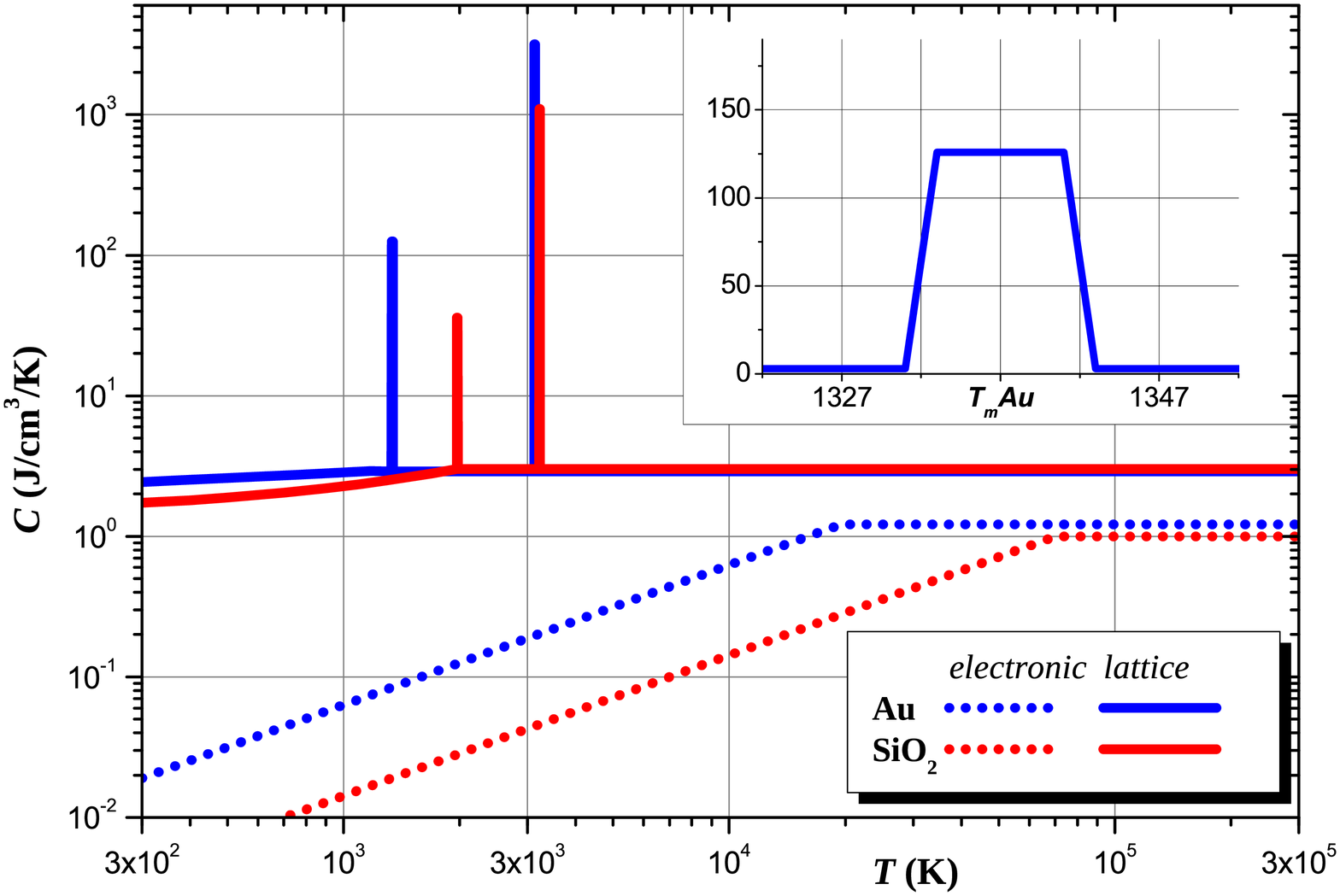}
		\caption{Electronic (C$_e$ in dotted lines) and atomic (C$_a$ in straight lines) specific heats of Au (blue) and SiO$_2$ (red) as a function of temperature}
		\label{fig:specheat}
  \end{minipage}\hfill
  \begin{minipage}{0.5\textwidth}   
		\centering	\includegraphics[scale=0.25]{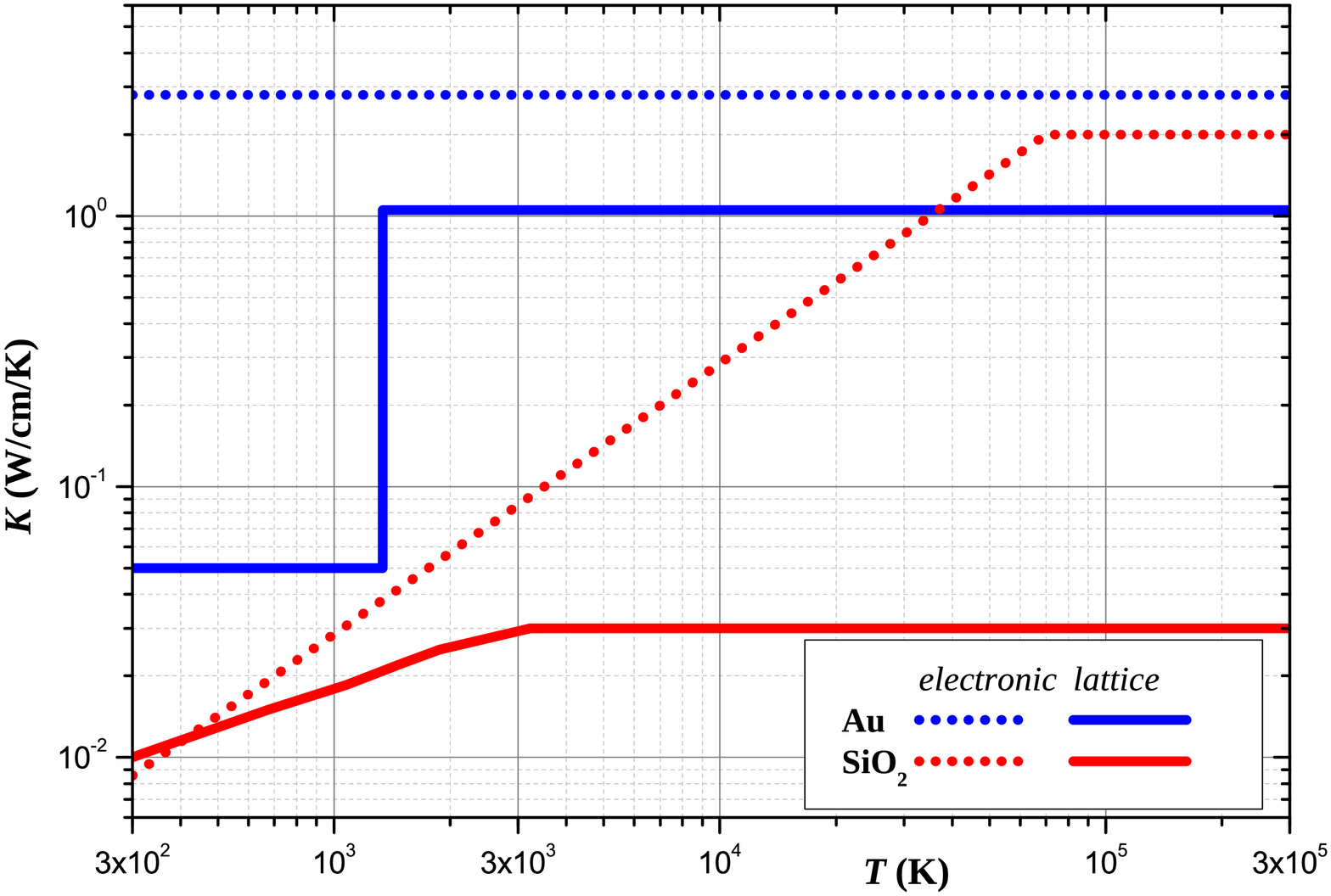}
		\caption{Electronic (C$_e$ in dotted lines) and atomic (C$_a$ in straight lines) thermal conductivities of Au (blue) and SiO$_2$ (red) as a function of temperature}
		\label{fig:thermocond}
  \end{minipage}\\
\end{figure}

The lattice specific heat $C_a$ and thermal conductivity $K_a$ values are a compilation of data from \cite{Meftah1994,CRCHandbook2006,Perry1984,Touloukian1970}. 
However, for gold, known as a noble metal, the main contribution to measured thermal conductivity comes from electronic part, especially at low temperature. So, in order to estimate the atomic thermal conductivity, we use the relation  $K_a=C_av_S\lambda/3$ , where $v_S$ is the sound speed and $\lambda$ is the phonon mean free path (of order of few lattice parameters), in the same way, $C_a$ is taken for classical statistical physics. The temperature dependence of $C_a$ and $K_a$ are plotted in  \fref{fig:specheat} and \fref{fig:thermocond} respectively (continuous lines: blue for Au and red for SiO$_2$).

The latent heat of  phase change, $Q_m$, is accounted using the following algorithm described in the case of solid-liquid change that occurs at $T=T_m$. For temperatures between $T_m+\Delta T$ and $T_m+\Delta T$, we modify the temperature dependence $C_a(T_a)$ by adding the peak-like function $C_m(T_a)$  to $C_a(T_a)$, so that $\int_{T_{m}-\Delta T}^{T_{m}+\Delta T} C_m(T) dT = Q_m$, $\Delta T = 5$K (See insert in \fref{fig:specheat}). Since the value of $C_a$ becomes high around melting point, the lattice temperature during melting remains practically constant. Similar modification is done also for vaporization account.

\begin{table}
\begin{tabular}{|c|c|c|}
\hline
	Material &$SiO_2$&$Au$\\
		\hline
	$\rho_S$[g/cm$^{3}$]&2.32&19.3\\
		\hline
	$\rho_L$[g/cm$^{3}$]&2.32&17.3\\
		\hline
	$T_{m}$[K]&1972&1337\\
		\hline
	$T_{v}$[K]&3223&3130\\
		\hline
	$Q_{m}$[J/g]&142&63.7\\
		\hline
	$Q_{v}$[J/g]&4715&1645\\

		\hline
		$g$[W/(cm$^{3}~$K)]&$1.25~ 10^{13}$ \cite{Awazu2008}&$2.3~ 10^{10}$ \cite{WANG1994}\\
		\hline
		$S_e$[keV/nm]&9.7&29.2\\
		\hline
		$S_n$[keV/nm]&0.025&0.113\\
		\hline
\end{tabular}
\caption{Physical parameters of SiO$_2$ and Au. }
	\label{tab:parameters}
\end{table}

\subsection{Electronic parameters} \label{elecparam}

We consider the electronic system in gold as free electron gas. At low temperatures $C_e(T_e)=\frac{\pi^2 k_B n_e}{2} \frac{T_e}{T_F}$ is linear with temperature according to \cite{Ashcroft1976}. Here $T_F=\frac{\hbar^2}{2m_e k_B}  (3\pi^2n_e)^{2/3}$ is the temperature corresponding to the Fermi energy, $\hbar$ and $k_B$ are Planck and Boltzmann constants, $m_e$ and $n_e$ are electron mass and number density, respectively. The high temperature classical value $C_e=\frac{3}{2}k_B n_e$ is valid for $T_e > \frac{3}{\pi^2}T_F$ . $C_e(T_e)$ is presented in \fref{fig:specheat} (blue dotted line).

Concerning the thermal conductivity, the relation $K_e=C_e D_e$ is used. The temperature dependence of the electronic diffusivity $D_e(T_e)$ was studied by Martynenko et al. \cite{Martynenko1983} and used here. It was shown that the value of $D_e$ decreases from $\sim10^2$ cm$^2$ s$^{-1}$ at $300$K down to $\sim 1$ cm$^2$s$^{-1}$ at $T_e\sim 10^4~$K$^{-1}$. So, we approximate $D_e(T_e)=300 \frac{D_{300}}{T_e}$ at $T_e<T_{lim}$, and $D_e=D_{min}$ at $T_e>T_{lim}$, with $D_{300}=150~$cm$^2$s$^{-1}$, $D_{min}=2~$cm$^2$s$^{-1}$, and $T_{lim}=300 \frac{D_{300}}{D_{min}} $. Hence the value $K_e$ is plotted versus $T_e$ in \fref{fig:thermocond} (blue dotted line).

For the dielectrics, the free electron gas model is valid only when the temperature exceeds the bandgap ($E_g = k_BT_g\approx9eV$ leading to $T_g\approx10^5$ K for SiO$_2$).  We use a high temperature ($T>T_g$) specific heat value of 1 J cm$^{-3}$K) \cite{Toulemonde1996}, which corresponds to hot electrons in SiO$_2$, whereas below this temperature, we refine the model and use a linear law $C_e(T_e)$. $C_e$ is plotted as function of $T_e$ in  \fref{fig:thermocond}. According to the same reference, the high temperature thermal conductivity of SiO$_2$ is $2$ W cm$^{-1}$ K$^{-1}$. We refine in the same way the temperature dependence $K_e(T_e)$ using a linear law. Both linear laws simply account for the fact that, in a first order approximation, the number of electrons involved in the thermal process is proportional to the temperature $T_e$ as long as $T_e \le T_g$. 
Thus, as a refinement to ref. \cite{Awazu2008}, we take this into account (See \fref{fig:thermocond}). 

\subsection{Electron-phonon coupling constant $g$}

The electron-phonon coupling coupling constant $g$ has been studied in supraconductors \cite{KAGANOV1957,ALLEN1987} and  the formulation of $g$ has been successfully extended to other type of materials (metals \cite{Dufourcondmatt1993,WANG1994,Dufour1996,Dufour1997}, semiconductors and insulators \cite{Toulemonde2000,Toulemonde2011}) . We use the values reported in these latter references. The value of $g$ and $K_e$ used here are consistent with the ones used by \cite{Toulemonde2000,Toulemonde2011} since the mean free path $\lambda$ ($\lambda^2=K_e / g$) remains unchanged between the present paper and those references.

Considering all the remarks above, we can now report in table \tref{tab:parameters} all the thermodynamical parameters used for the calculations.

\section{Results and discussion}

\subsection{Comparison with previous calculations: cylindrical particles}

\begin{figure}
  \begin{minipage}{0.5\textwidth}
      \centering {\includegraphics[scale=0.6]{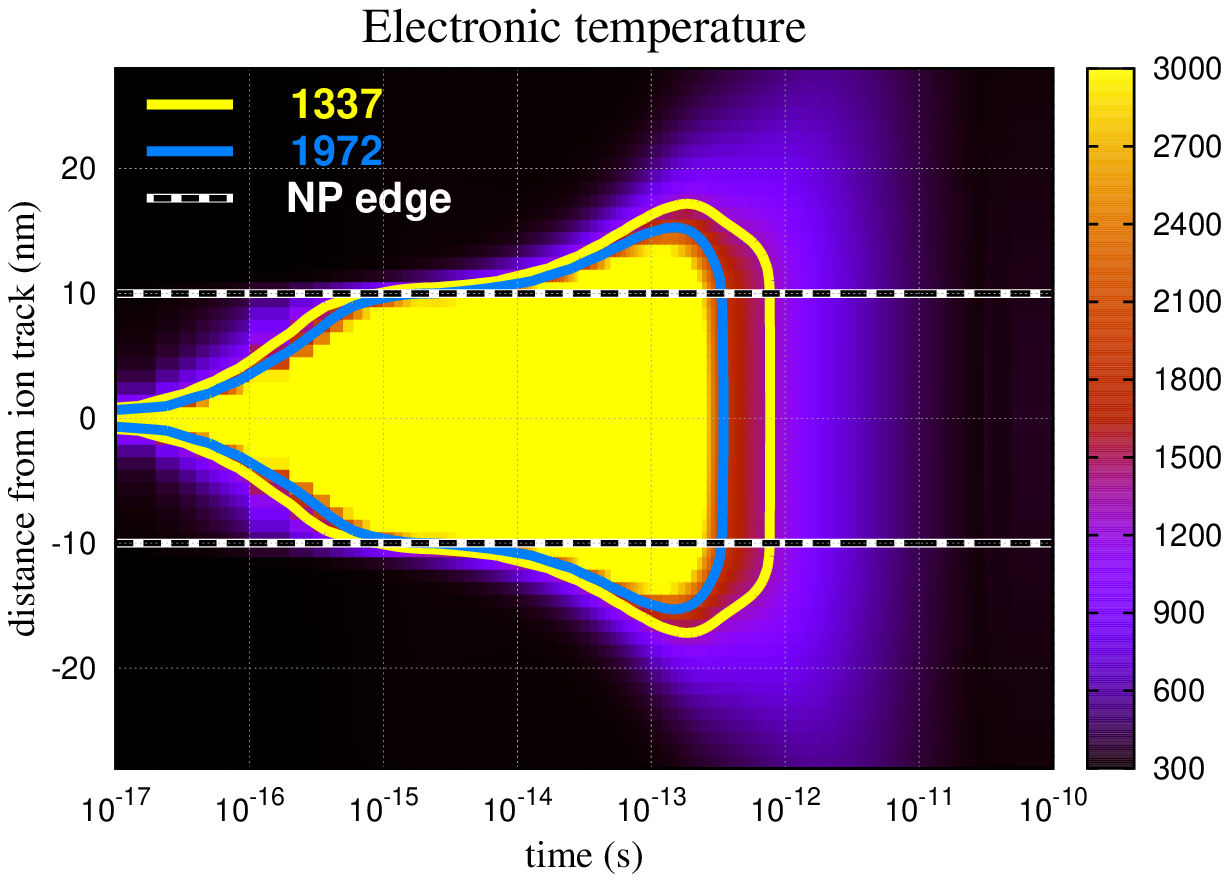}\\a)}
   \end{minipage}\hfill
   \begin{minipage}{0.5\textwidth}   
      \centering {\includegraphics[scale=0.6]{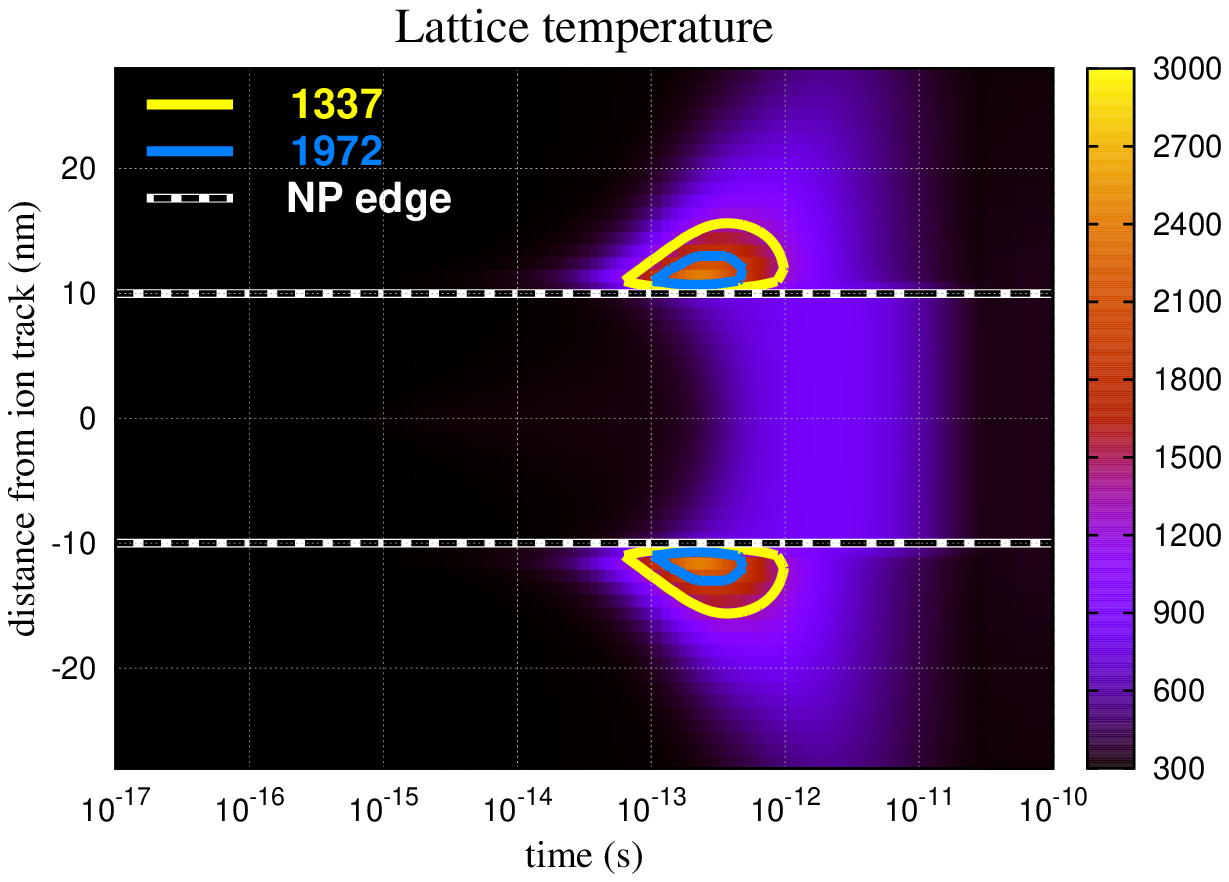}\\b)}
   \end{minipage}\\
 \caption{Cylindrical gold nanoparticle (diameter: $20nm$). Evolution of the electronic and atomic temperatures $T_e(r,t)$ and  $T_a(r,t)$ in Kelvin as a function of radial distance $r$ from the ion path and time $t$.The thermodynamic parameters are those from Awazu et al. \cite{Awazu2008} } %
 \label{cyl20Aw}
\end{figure}

In this section, we make the link between the present study and the work previously published by Awazu et al. \cite{Awazu2008}. In order to confirm the accuracy of the present 3D version of the thermal spike code, we have performed simulations first in bulk gold and silica, and then in the case of  gold cylindrical nanoparticles with 10 nm and 20 nm diameter embedded in silica with thermodynamic parameters used in \cite{Awazu2008}. The results are perfectly consistent with \cite{Awazu2008} regarding the maximal temperatures reached as well as the heating and cooling times. We plot both electronic and atomic temperature evolutions as a function of time (abscissa) and radial distance to the ion path (ordinate) for 20 nm particle (\fref{cyl20Aw}). Like for simulation done in \cite{Awazu2008} the gold nanoparticle do not melt. This is due to the fact that electronic heat distributed over NP spreads out and is transferred to silica atoms before 
 $10^{-12}$ s, while this time is too short for the gold lattice to heat up due to the low $g$-factor. 

However, elongation of such nanoparticles \cite{Awazu2008}, and even larger ones (40 nm \cite{Awazu2009Nano}) was observed experimentally.

\begin{figure}
  \begin{minipage}{0.5\textwidth}
      \centering {\includegraphics[scale=0.6]{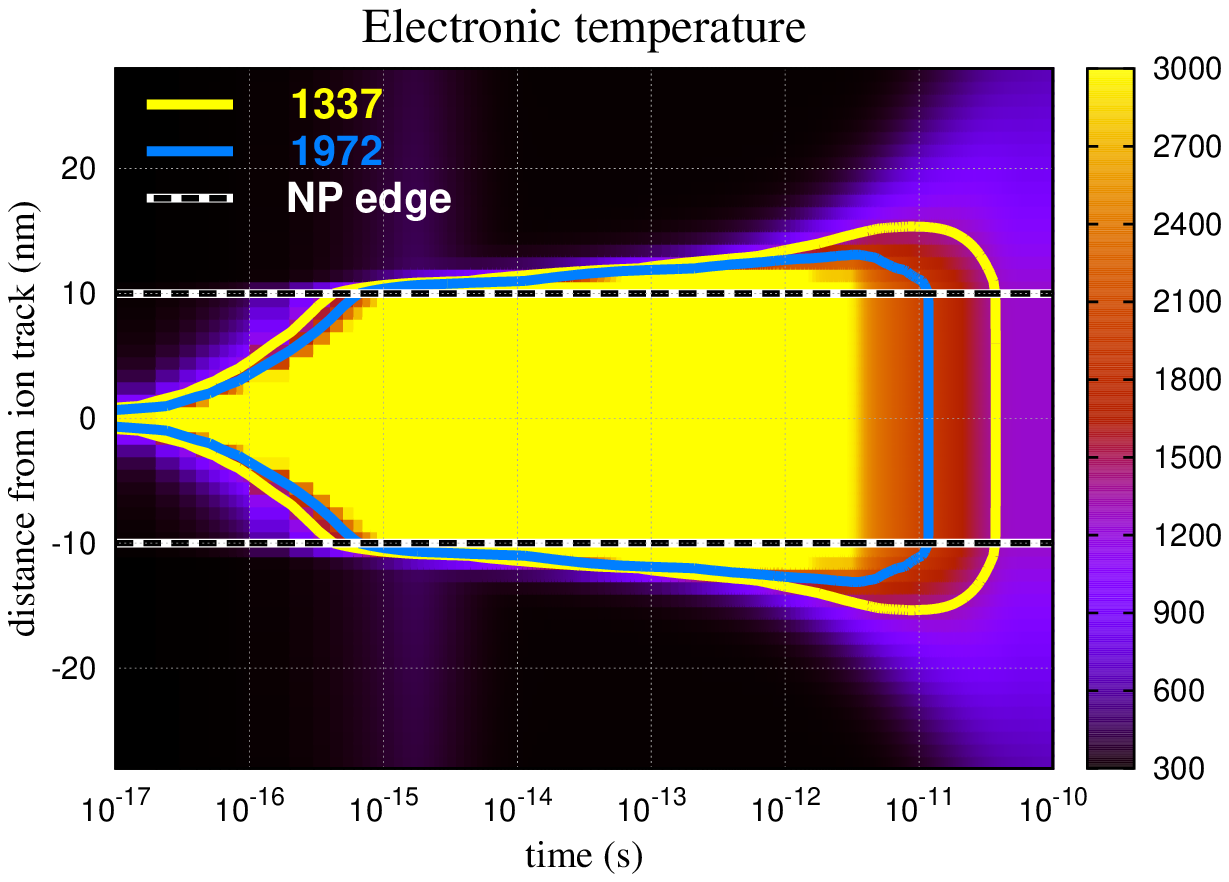}\\a)}
   \end{minipage}\hfill
   \begin{minipage}{0.5\textwidth}   
      \centering {\includegraphics[scale=0.6]{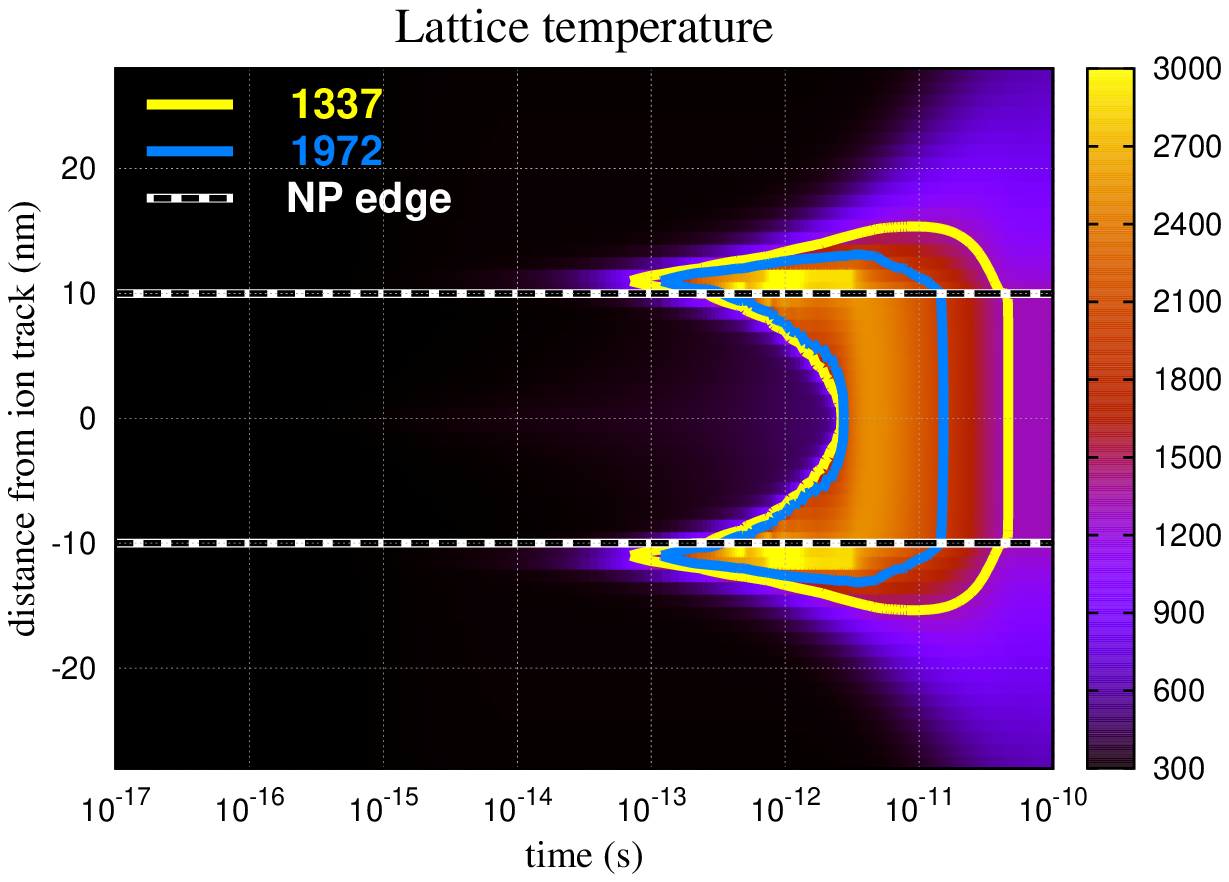}\\b)}
   \end{minipage}\\
 \caption{Cylindrical gold nanoparticle (diameter: $20nm$). Evolution of the electronic and atomic temperatures $T_e(r,t)$ and  $T_a(r,t)$ in Kelvin as a function of radial distance $r$ from the ion path and time $t$. The thermodynamic parameters and the thermal conductivities $K_{e_{SiO_2}}$ and   $K_{a_{Au}}$ are those of the present work.}
 \label{cyl20Kh}
\end{figure}

We show the very different behaviour (see \fref{cyl20Kh}) if we use the SiO$_2$ $K_e(T_e)$  value (\fref{fig:thermocond}) as previously discussed (see sect. \sref{elecparam}). Up to $t\approx2~10^{-14}$ s, the electronic temperature for both cases is similar, but for greater times, (\fref{cyl20Aw}a), the temperature increase spreads out of the nanoparticle, while hot electrons are confined within the NP (\fref{cyl20Kh}a) so that $T_e$ remains higher than $3000$K for times as long as $3~ 10^{-12}$ s inside the NP as well as in the thin SiO$_2$/NP interface layer. Due to high $g$-factor electronic heat is efficiently transferred to the silica atoms so that the SiO$_2$ lattice gets also hot in this interface layer. Finally, due to lattice thermal diffusion, the lattice temperature of metal nanoparticle rises. The heating of nanoparticle atoms comes from the outer layers toward the core until complete NP melting(see \fref{cyl20Kh}b). This process takes time from $10^{-13}$ s to $3~10^{-12}$ s, then both electronic and lattice subsystems are in equilibrium and cool down until the NP is solidified at about $5~10^{-11}$ s.

\subsection{Spherical gold nanoparticles}

Actually, the simulations from previous section, both ours and those described in \cite{Awazu2008}, were performed in 2D-plane perpendicular to swift heavy ion track which avoids investigations as a function of depth ($z$-coordinate).  Now, we deal with spherical particles embedded in a matrix so that the third coordinate is required since the incoming ion first penetrates the matrix and then the particle. We again compare the simulations resulting from the parameter set used by Awazu et al.\cite{Awazu2008} and the present one for two particle sizes.

\subsubsection{$20nm$ gold nanoparticle: role of the thermal conductivity at the SiO$_2$ interface}

\begin{figure}[hb]
  \begin{minipage}{0.5\textwidth}
      \centering {\includegraphics[scale=0.6]{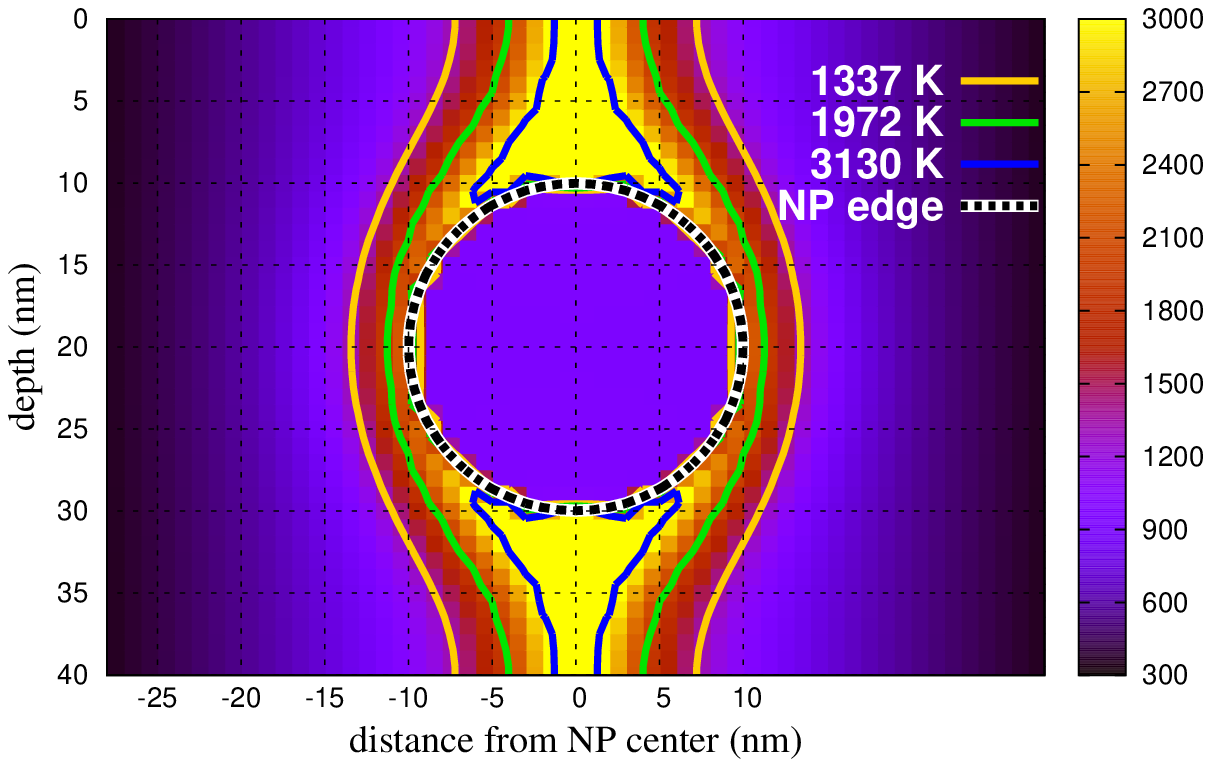}\\a)}
   \end{minipage}\hfill
   \begin{minipage}{0.5\textwidth}   
      \centering {\includegraphics[scale=0.6]{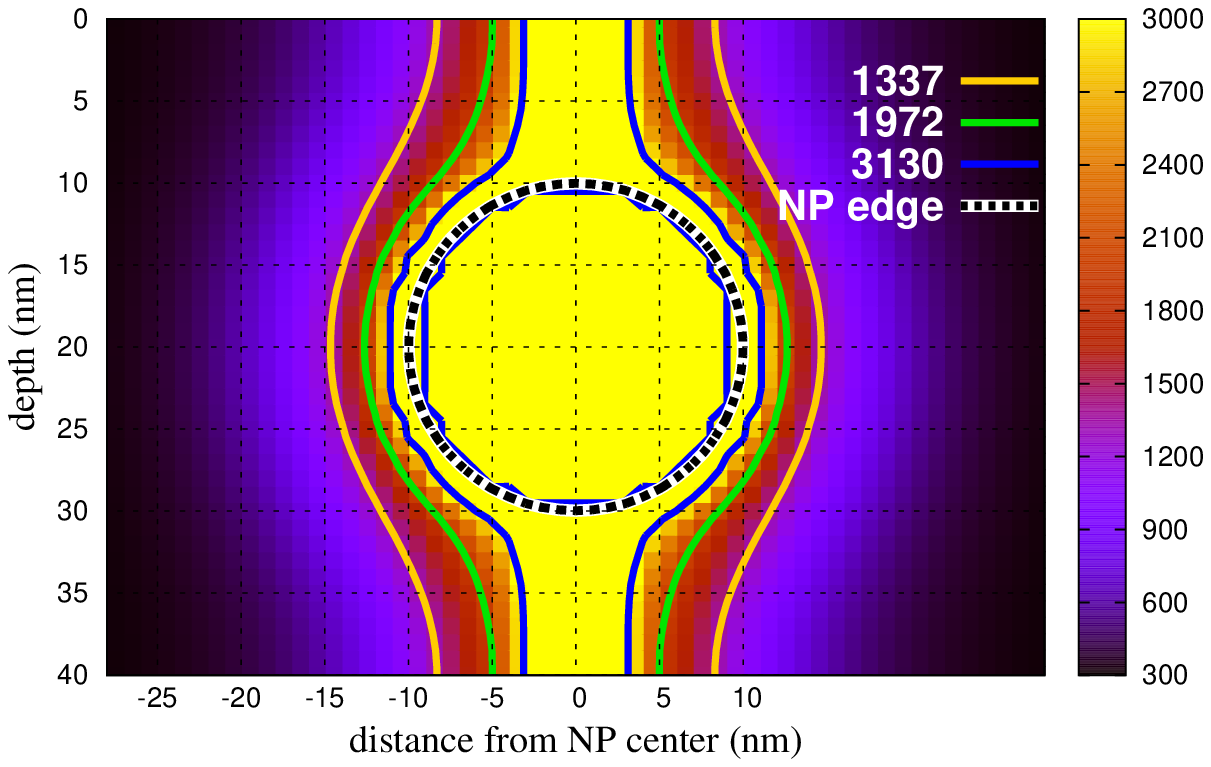}\\b)}
   \end{minipage}\\
 \caption{Maximum lattice temperature reached around a gold nanoparticle (diameter = 20 nm). Physical parameters: a) Awazu et al. \cite{Awazu2008}, b) present paper. Contour lines correspond to: $Au$ melting temperature (1337K), $SiO_2$ melting temperature (1972K), $Au$ vaporizing temperature (3130K). The dotted blue line is the boundary of the metallic particle. }
 \label{sph20}
\end{figure}

In \fref{sph20} the maximum lattice temperature reached in 20 nm diameter spherical particle is shown. One can see that in case of Awazu's parameters (left hand side), only silica regions close to the ion path or NP surface are hot, since they are provided with high electronic temperature at early stage. The outside area and particle itself remain relatively cold.
In contrary, if we account for the thermal conductivity (see sect. \sref{elecparam}) (right hand side), we observe very hot track in silica, and completely molten nanoparticle with vaporized outer layer, while outer space is almost at initial temperature. These results are similar to previous section, but the extra heat comes to polar NP regions from neigbouring silica.

\subsubsection{$40nm$ gold nanoparticle: influence of the gold lattice thermal conductivity}

\begin{figure}[ht]
  \begin{minipage}{0.5\textwidth}
      \centering {\includegraphics[scale=0.6]{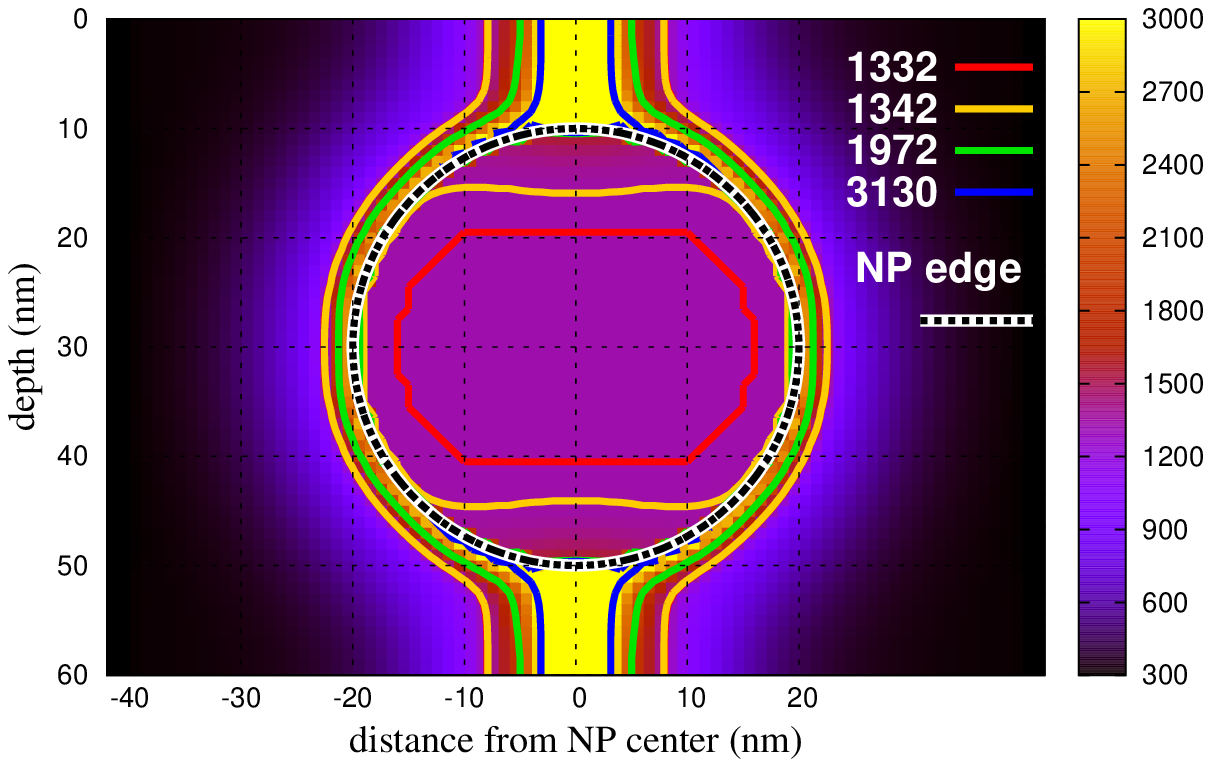}\\a)}
   \end{minipage}\hfill
   \begin{minipage}{0.5\textwidth}   
      \centering {\includegraphics[scale=0.6]{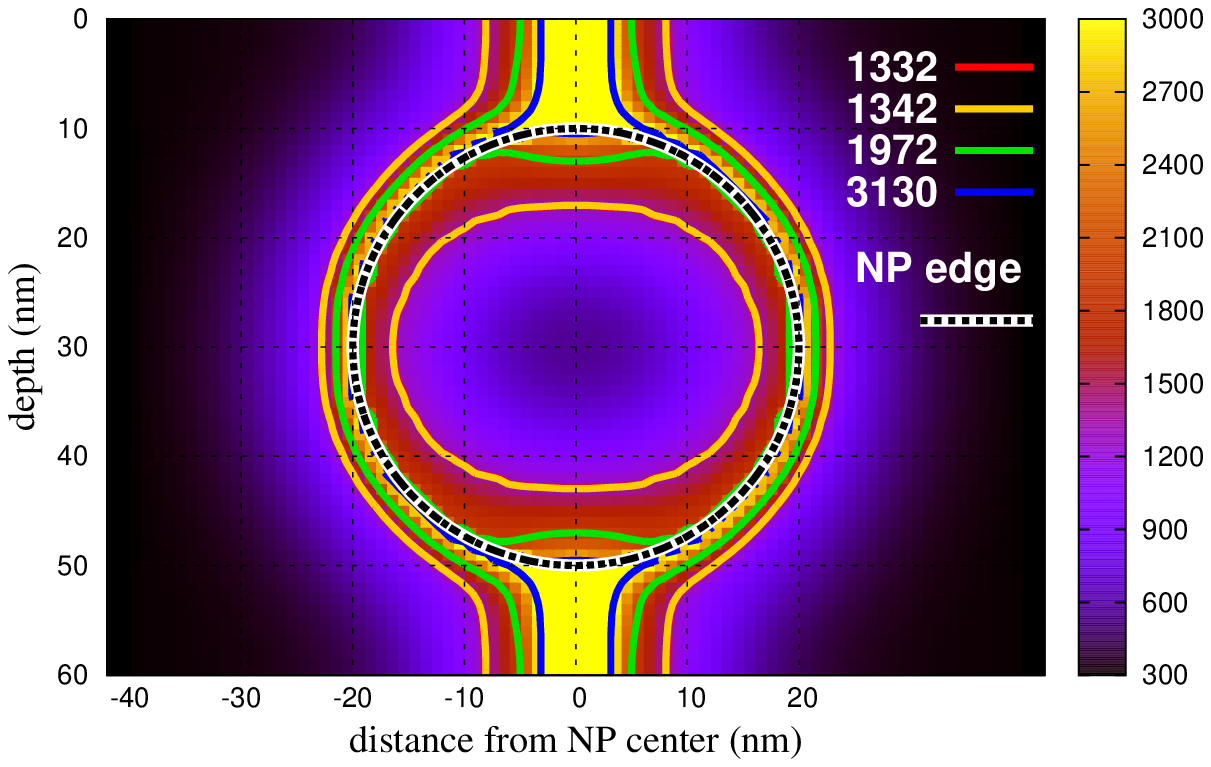}\\b)}
   \end{minipage}\\
 \caption{Maximum lattice temperature reached in a  gold particle ($d = 40 nm$), using different thermal conductivities ($K_a$): a) $K_a$ taken from Awazu \cite{Awazu2008}, b) our proposal (see sect. \sref{lattparam}). Contour lines correspond to the melting points: 1332K - $T_{m-}$(gold), 1342K - $T_{m+}$ (gold), 1972K - $T_m$ (SiO$_2$). 3130K is the vaporizing temperature of  gold.}
 \label{sph40}
\end{figure}

Another peculiarity can be found for larger particles which are not completely molten ( \fref{sph40}). 
Concerning the proper choice of lattice thermal conductivity of a metal, as we mentioned above, the measured values are not adequate since they mostly represent the electronic part of thermal conductivity. The molten phase on the plots is presented with double curves, at $T_{m}\pm \Delta T$ (1332K and 1342K). It can be seen that, with measured values of $K_a$ (\fref{sph40}a), an extended region reaches melting point ($T_{m-}$ contour), though only a small fraction has surpassed the melting latent heat ($T_{m+}$ contour). With the correct $K_a$ value (\fref{sph40}b), both contours coincide and the volume of complete melting is much larger, while the core temperature remains lower than on \fref{sph40}a.

\subsection{Melting and vaporization of nanoparticles with different sizes}

\begin{figure}[ht]
  \begin{minipage}{0.5\textwidth}
		\centering\includegraphics[scale=0.25]{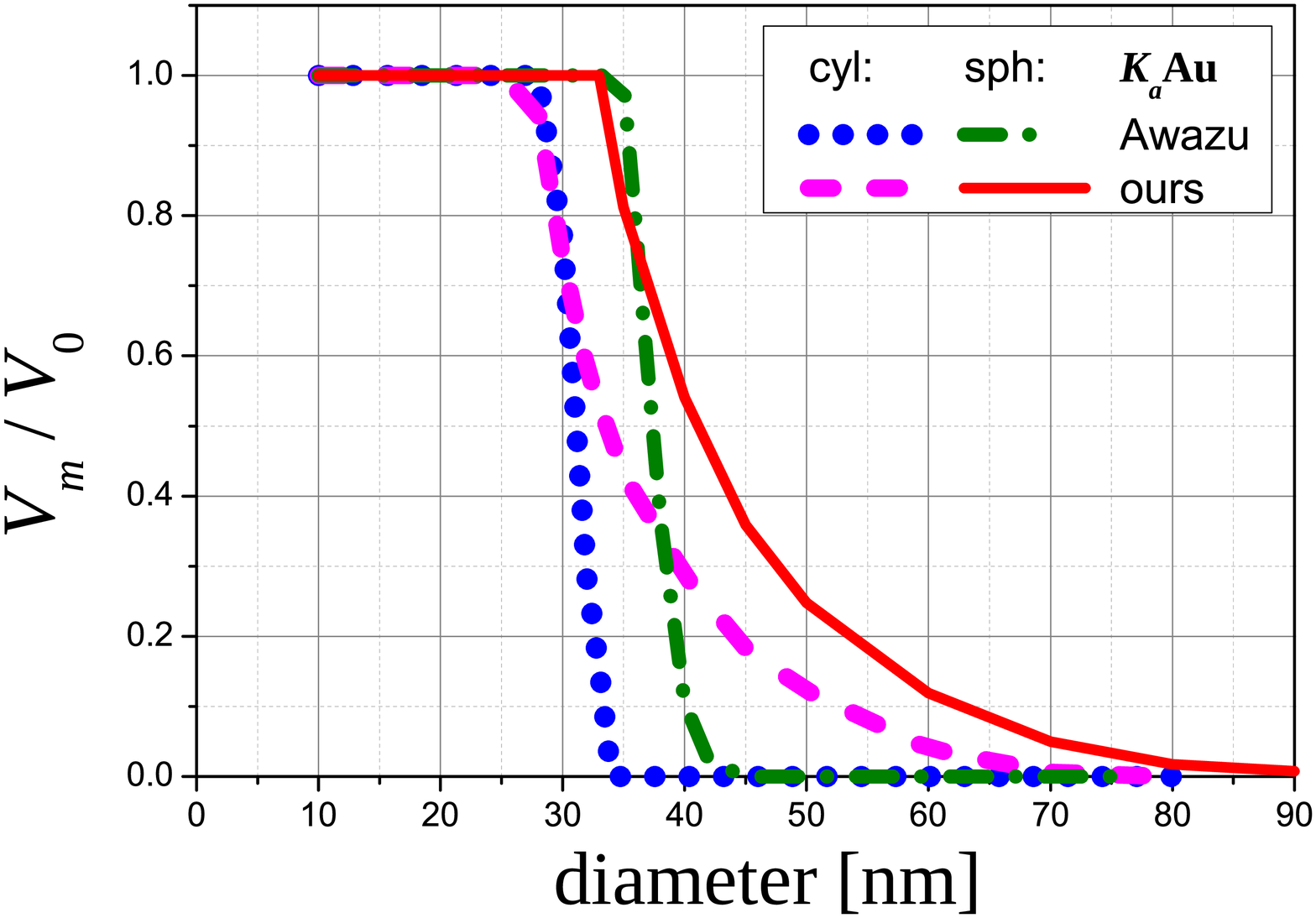}
  \end{minipage}\hfill
  \begin{minipage}{0.5\textwidth}   
		\centering\includegraphics[scale=0.25]{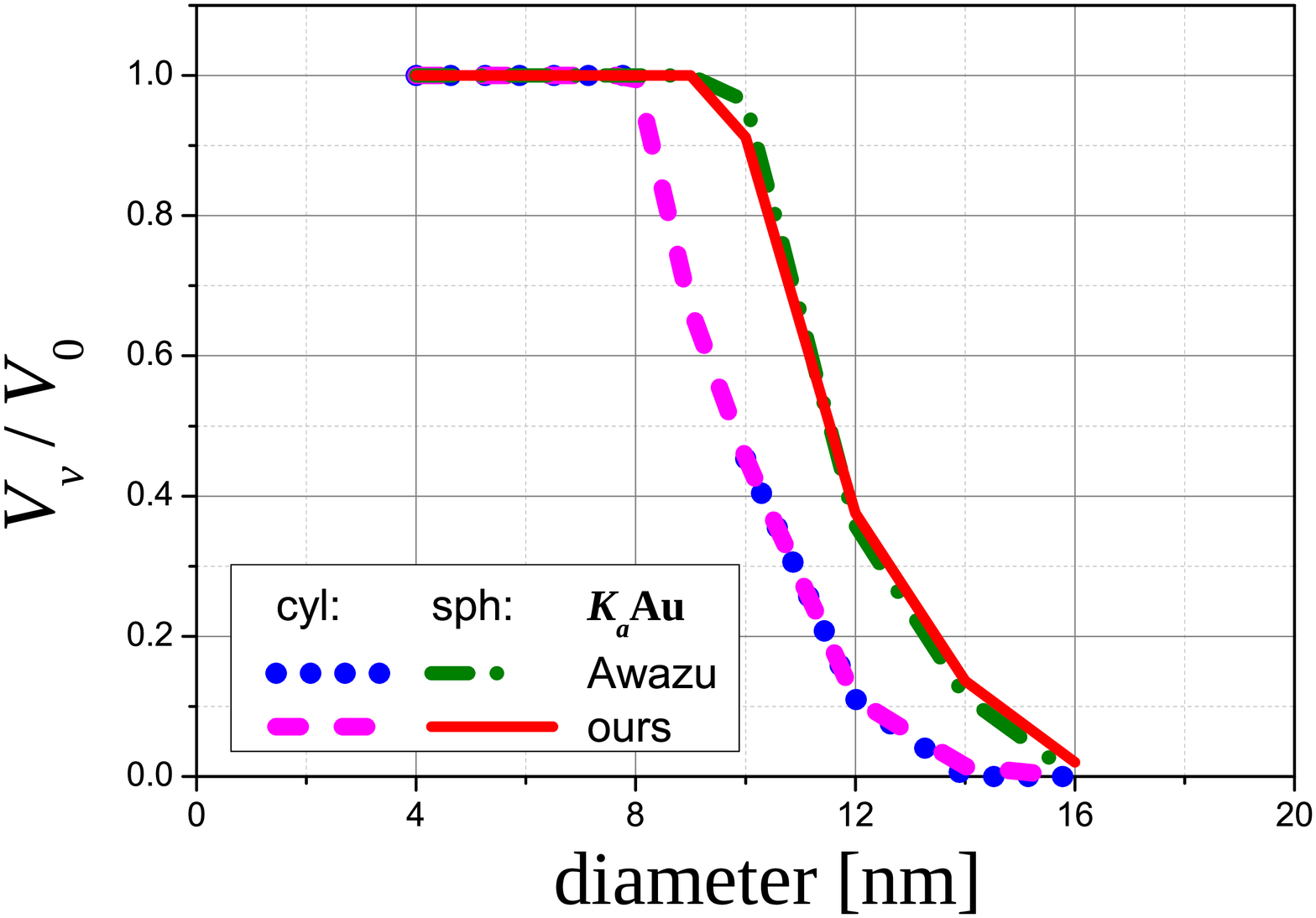}
  \end{minipage}\\
	\caption{Molten (a) and vaporized (b) fraction vs. particle diameter}
	\label{fig:Vmv_size}
\end{figure}

The volumic molten and vaporized fraction vs. initial particle diameter is shown on \fref{fig:Vmv_size} for both spherical and cylindrical particles with two kinds of lattice thermal conductivity of gold and interfacial electronic thermal conductivity. As we showed above, if we do not account for the interfacial region, melting is not observed  even for 20nm particles.

The relative molten and vaporized volume is systematically higher for 3D-spherical particles in comparison with cylindrical ones. We can suppose that the energy deposited in the nanoparticle by the incident ion is then confined and evenly distributed over the whole volume of the particle. The value $\varepsilon=(S_e l)/(V N_a)$ gives the energy density per atom received from a swift heavy ion. Here $l$ is the ion path inside the particle, $V$ and $N_a$ are particle volume and atomic concentration.
For a cylinder with a radius $R$, $\varepsilon_{cyl}=\frac{S_e}{\pi R^2}$, while for a sphere $\varepsilon_{sph}=\frac {3}{2}\frac{S_e }{\pi R^2}$ (in case of central ion hit). Naturally, molten (and vaporized) fraction is higher for spherical NP than for cylindrical one of the same diameter. Also, the heat exchange between metal and silica around ion spot plays important role.

Complete melting is observed up to 30nm diameter. Then, if ion energy is not high enough to melt the whole particle, then with experimental (high) lattice thermal conductivity of gold the molten fraction quickly drops to zero, while using $K_a$ derived from the equation in section \sref{lattparam}, the molten fraction slowly decreases, remaining non-zero up to (70-80)nm.

As for vaporization, there is no difference between $K_a Au$ from \cite{Awazu2008} and present work, since above melting point both parameters are the same. The particles with diameter from 10 to 15 nm are partially vaporized, while below 10nm complete vaporization is observed. Such small particles undergo dissolution and provide precipitated material for larger particles.   

\subsection{Full 3D: effect of off-axis impact}

\subsubsection{Temperature distribution}
\begin{figure}
   \begin{minipage}[hb]{0.5\textwidth}   
      \centering {\includegraphics[scale=0.3]{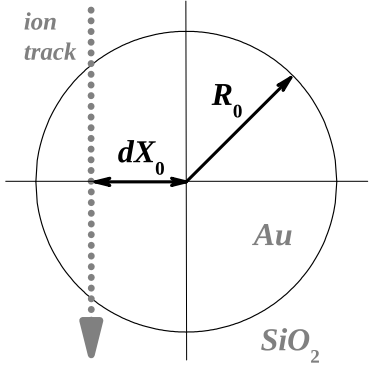}}
   \end{minipage}\\
 \caption{Off-axis impact layout.}
 \label{figschema2}
\end{figure}

Using the 3D capabilities to treat any problem without any symmetry, we focus on the effect of an ion off-axis impact (\fref{figschema2}). We performed  simulation in NP diameter range from 10nm to 80nm at different offset values: $\Delta=dX_0/R_0=n/4$, $n$ varied from 1 to 6. 

\begin{figure}
  \begin{minipage}[b]{0.5\textwidth}
      \centering \includegraphics[scale=0.6]{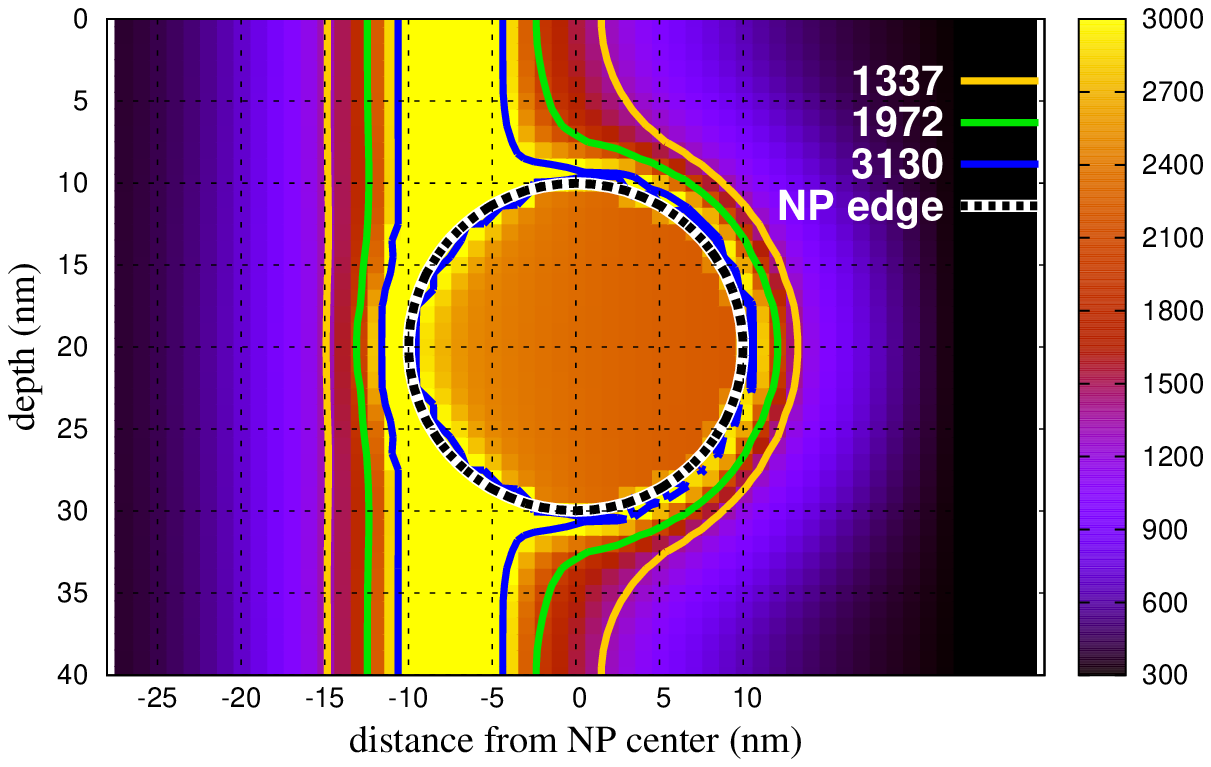}
   \end{minipage}\hfill
   \begin{minipage}[b]{0.5\textwidth}   
      \centering \includegraphics[scale=0.6]{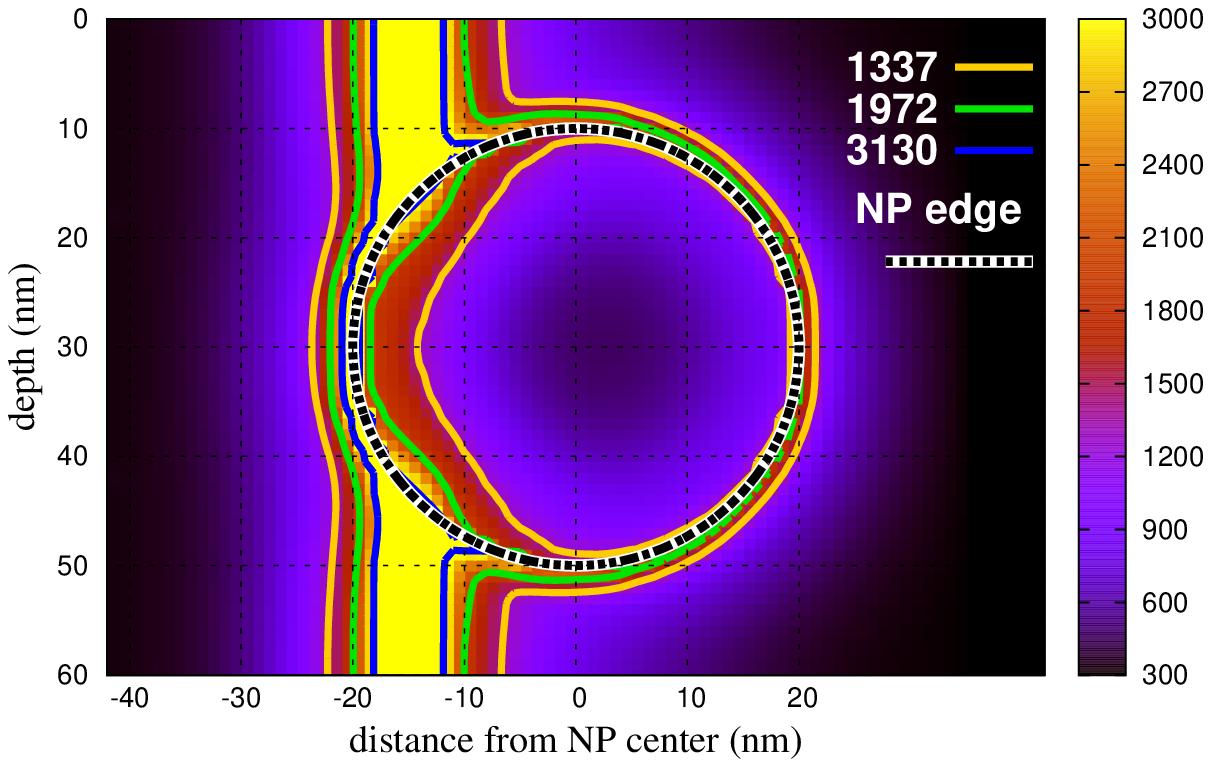}
   \end{minipage}\\
  \begin{minipage}[b]{0.5\textwidth}
      \centering \includegraphics[scale=0.6]{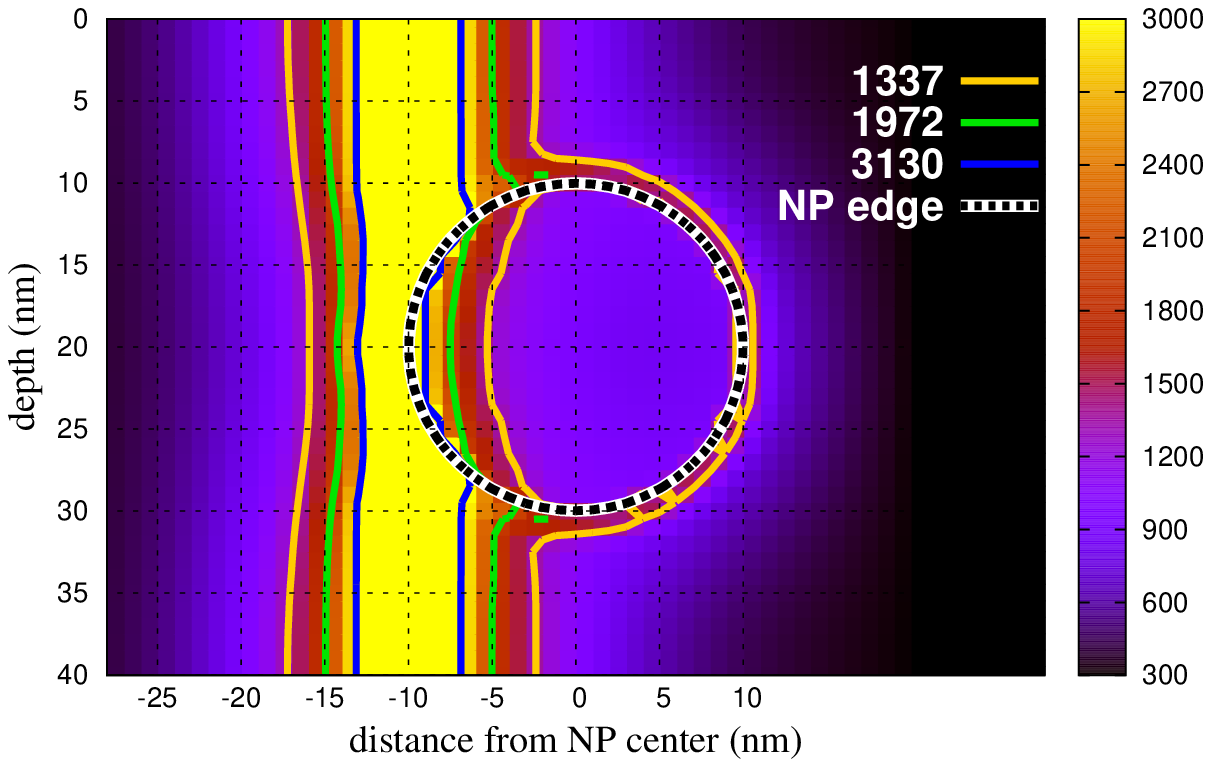}
   \end{minipage}\hfill
   \begin{minipage}[b]{0.5\textwidth}   
      \centering \includegraphics[scale=0.6]{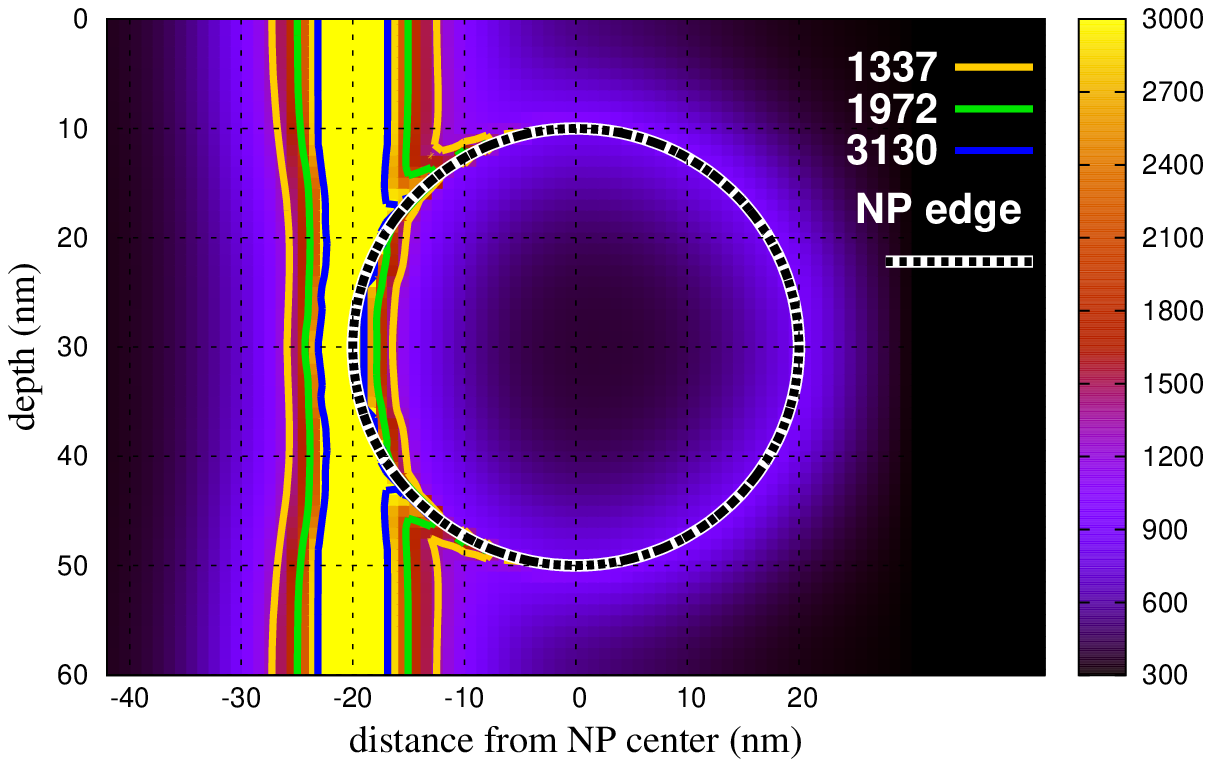}
   \end{minipage}\\
  \begin{minipage}[b]{0.5\textwidth}
      \centering {\includegraphics[scale=0.6]{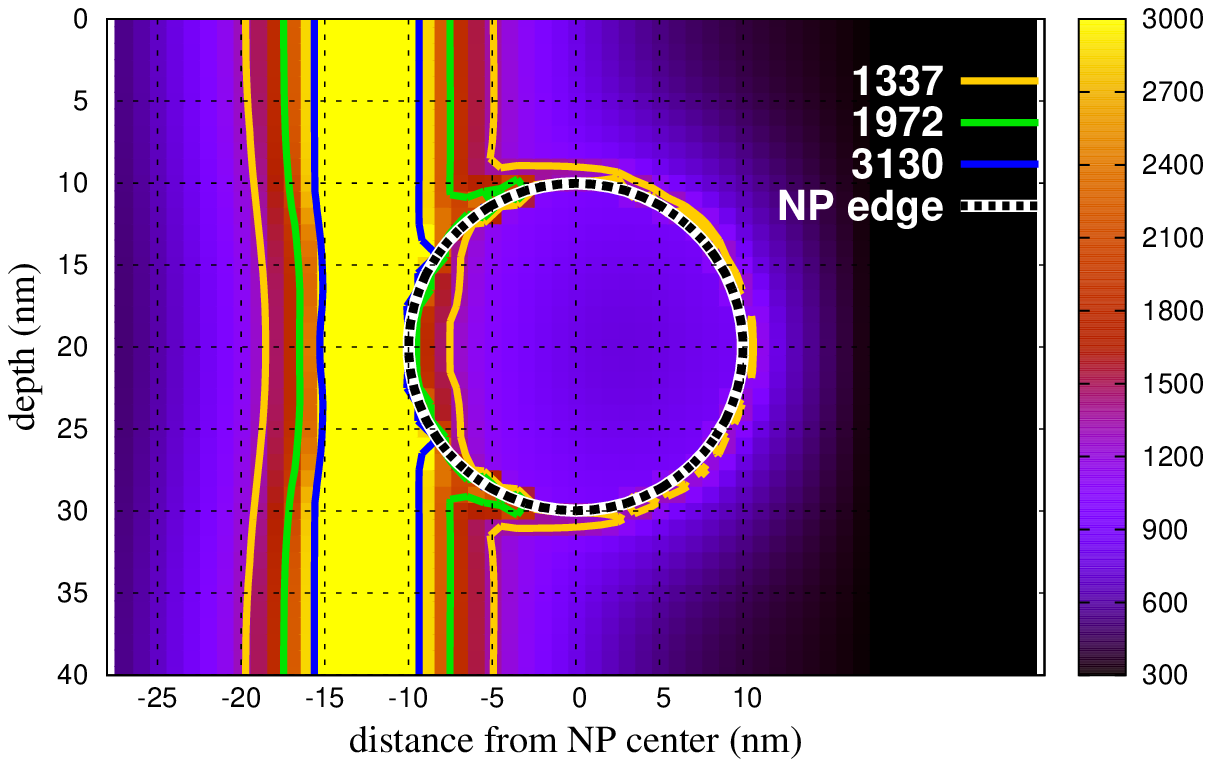}\\a)}
   \end{minipage}\hfill
   \begin{minipage}[b]{0.5\textwidth}   
      \centering {\includegraphics[scale=0.6]{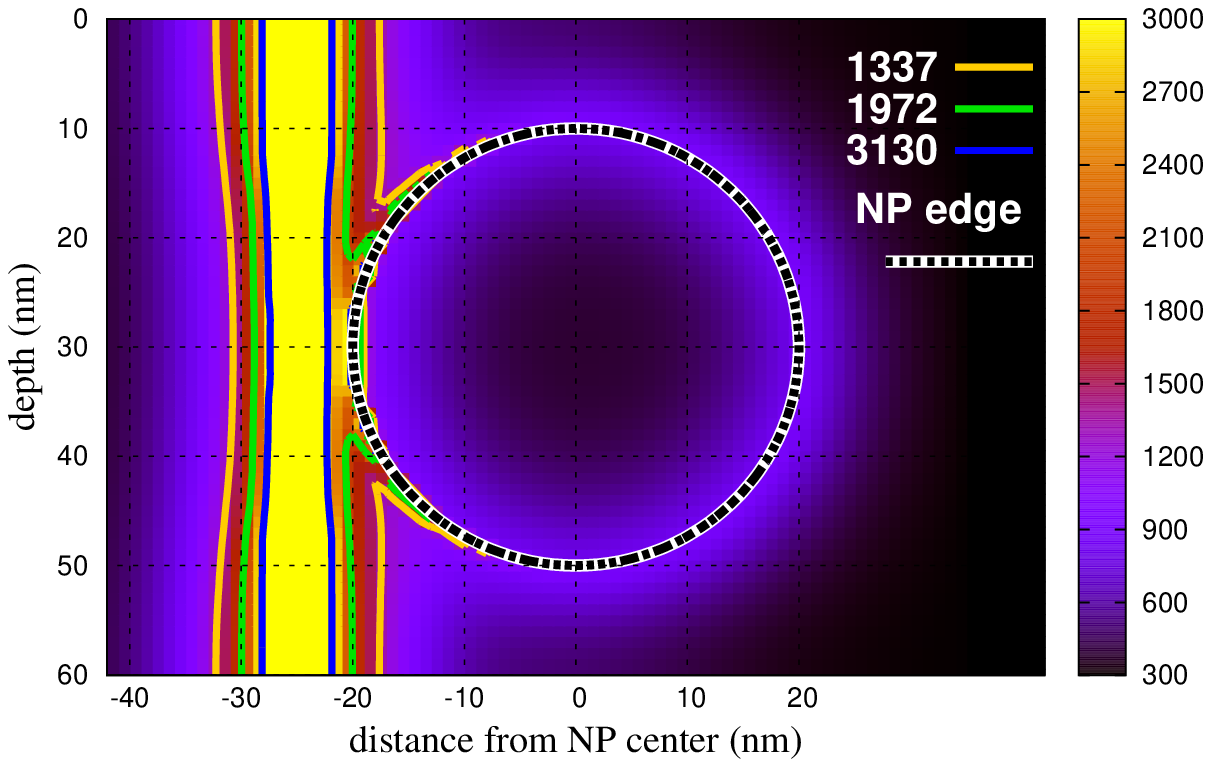}\\b)}
   \end{minipage}\\
 \caption{Maximum lattice temperature as a function of the ion path distance $d$ ($d=R . \Delta$) from the center of a gold nanoparticle of radius $R$. Left column: particle diameter $D=20 nm$; from top to bottom: $\Delta=\frac{3}{4},1,\frac{5}{4}$. Right column: particle diameter $D=40nm$; from top to bottom: $\Delta=\frac{3}{4},1,\frac{5}{4}$}
 \label{asym20-40}
\end{figure}

The most interesting to see what happens when the ion hits close to the edge of NP. The corresponding plots are shown in \fref{asym20-40} for 20nm and 40nm nanoparticles and $\Delta=\frac{3}{4},1,\frac{5}{4}$ (the central impacts for them are shown on \fref{sph20}b,\fref{sph40}b). Two remarks arise from the results: i) in non central impacts, the melting of the particle still starts from the surface, ii) the lattice temperature significantly increases only if the ion path intercepts the particle. 

This last point leads to the fact that, for such particles, a minimum fluence $\Phi_{min}$ of the order of $\frac{1}{\pi R^2}$ ions cm$^{-2}$ is required to transform all the particles i.e. to obtain a fusion at the particle surface.  $\Phi_{min}= 8~10^{10}$ cm$^{-2}$ (respectively $3.2~10^{11}$cm$^{-2}$) for particle diameter of $40$nm (respectively $20$nm) Refering to \cite{Ridgway2011}, the minimum fluences cited giving place to particle elongation is of the order of some $10^{11}$ ions cm$^{-2}$ that means that a given particle has been hit by several ions which by statistical effect act as if the mean impact was centered so that the elongation eventually occurs along the irradiation beam direction.

\subsubsection{Energy density and molten volume}

\begin{figure}[ht]
  \begin{minipage}{0.5\textwidth}
      \centering {\includegraphics[scale=0.25]{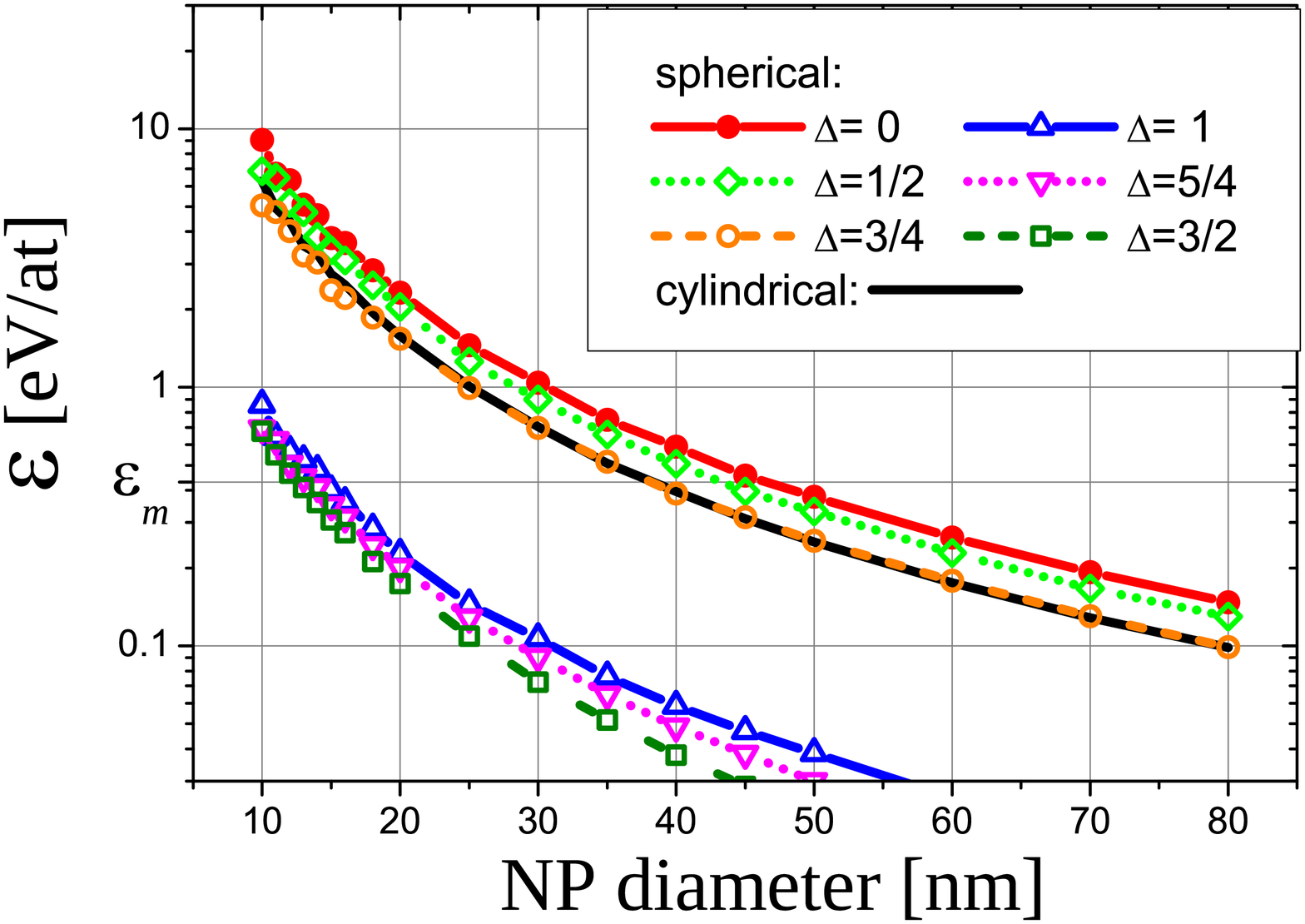}\\a)}
   \end{minipage}\hfill
   \begin{minipage}{0.5\textwidth}
      \centering {\includegraphics[scale=0.25]{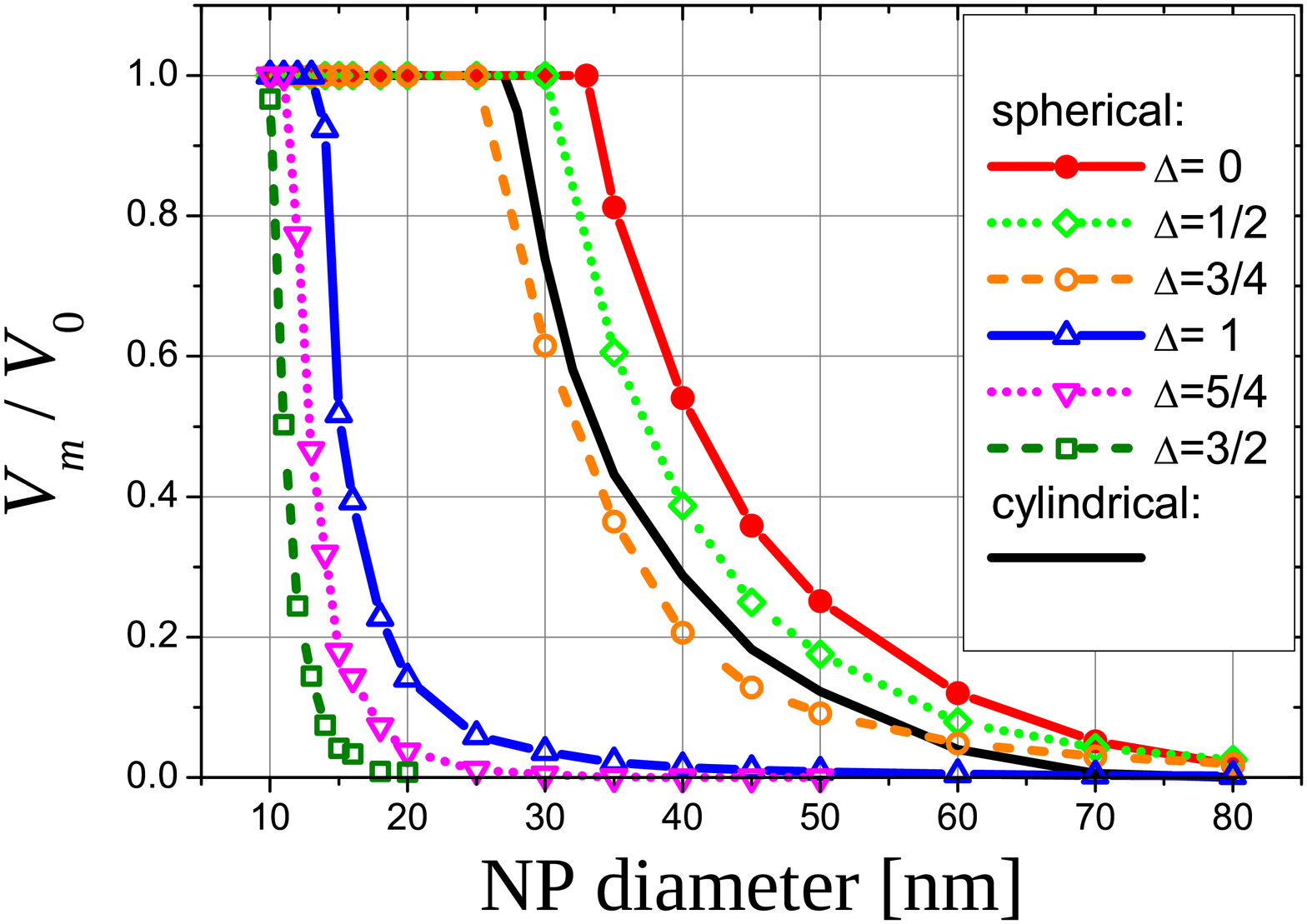}\\b)}
   \end{minipage}\\
 \caption{Deposited energy density (a) and molten volume fraction (b) versus the gold nanoparticle diameter for different values of the offset $\Delta$ (see legend \fref{asym20-40} for the definition of $\Delta$)}
 \label{EVm_d}
\end{figure}

\fref{EVm_d} demonstrates the influence of the nanoparticle diameter on the deposited energy density ($\varepsilon$) and the molten volume fraction ($V_m/V_0$) for spherical and cylindrical particles. $\varepsilon$ and $V_m$ are plotted as a function of the ion path offset $\Delta$. First, there is a difference depending on whether the incident ion enters  NP ($\Delta<1$) or not ($\Delta\geq1$). In the latter case, only a small amount of energy is deposited due to secondary $\delta-$electrons penetrating inside the particle. However this is enough to melt nanoparticles smaller than $\sim$20nm in radius. Then, comparing the deposited energy density for spherical $\varepsilon_{sph}=\frac {3}{2}\frac{S_e }{\pi R^2}\sqrt{1-\Delta^2}$ and cylindrical particles $\varepsilon_{cyl}=\frac{S_e }{\pi R^2}$, one can see that for $\Delta=\sqrt{5}/3\approx 3/4$, $\varepsilon_{sph}=\varepsilon_{cyl}$. The $\varepsilon(d)$ curves really coincide for this latter case (\fref{EVm_d}a), but due to electronic heat leakage from the particle to the surrounding silica in the ion spot area less spherical particle is molten in diameter range (25-60)nm. However, starting from 60nm diameter, all spherical particles intercepted by ion ($\Delta<1$) demonstrate stable molten fraction above $\sim2\%$, because heated  particle-SiO$_2$ interface keeps the temperature high enough to melt the metal.

We estimate the energy density needed to melt the whole particle as $\varepsilon_m=(\int_{T_0}^{T_{m}} C_a dT_a +\rho_S Q_m)/N_a$. For gold $\varepsilon_m=0.43$ eV/at. However, comparing \fref{EVm_d}a and b, one can see that the energy necessary for complete NP melting is approximately twice as large as  $\varepsilon_m$ (from $0.8$ to $1$ eV/atom). This is due to heat leakage outisde the particle and can be taken into account for estimation with other particles and ions. The corresponding curves for cylindrical particles do not depend on the value of the impact offset $\Delta$ provided that $0<\Delta<1$ (i.e. when the ion strikes the particle).

\section{Conclusion}

We presented the first real 3D implementation of thermal spike model devoted to the simulation of swift heavy ion interaction with any kind of composite structures. 
A quantitative description of the ion shaping of metallic nanoparticles embedded in amorphous matrix can be done provided that the thermodynamical parameters follow the characteristics described hereafter. The target electrons are considered as a quasi free electron gas with classical specific heat and thermal diffusivity, whereas the target lattice is characterized with thermal coefficients based on statistical physics. The measured values cannot be used for metals since the thermal properties essentially depend on the electrons so that none of the subsystems (electrons and lattice atoms) is well described. For example, in metals, the electronic thermal conductivity is much higher than the lattice thermal conductivity. Using the 3D approach, we can make the distinction between cylindrical and spherical particles and show the influence of energy exchange on the metal/silica interface around the ion spot. Using the metallic (nanoparticle) and dielectric (matrix) thermal properties, it is shown that the heating of nanoparticle starts from the outer layer toward the core. This provides conditions for partial melting when there is not enough energy to melt the whole nanoparticle.
Finally we demonstrate the influence of the impact parameter (i.e. radial distance between the ion path and the spherical nanoparticle center): $i$)a  large particle (radius $ 60$ nm) partially melts in the case of an ion path nearly  tangent to the particle surface due to large NP/matrix interfacial area; $ii$) a small particle ($ 60$ nm ) can be molten even if the ion does not penetrate it due to heat transfer from molten silica track of about 10 nm diameter.
The simulations made in present work essentially increase the limits of probability to melt metal nanoparticles irradiated with swift heavy ions and, thus, provide background for understanding of ion shaping effect.
\ack
The present work is partially supported by the French National Research Agency (ANR) in the framework of SHAMAN project (ANR-09-BLAN-0334) which involves two laboratories of IRAMIS section of CEA (France) ( LSI-Laboratoire des Solides Irradi\'es, Palaiseau (France) and CIMAP, Caen (France)),as well as the Laboratoire de Physique des Nanostructures (LPN), Marcoussis (France) and the Laboratoire de Physique de la Mati\`ere Condens\'ee (LPMC-CNRS UMR7643)  Ecole Polytechnique, Palaiseau.

\newpage
\section*{References}
\bibliography{jpdvladchris}

\end{document}